\documentclass[conference]{IEEEtran}
\IEEEoverridecommandlockouts
\usepackage{cite}
\usepackage{amsmath,amssymb,amsfonts}
\usepackage{algorithmic}
\usepackage{graphicx}
\usepackage{textcomp}
\usepackage{xcolor}
\def\BibTeX{{\rm B\kern-.05em{\sc i\kern-.025em b}\kern-.08em
    T\kern-.1667em\lower.7ex\hbox{E}\kern-.125emX}}

\usepackage{acronym}
\usepackage{hyperref}
\usepackage{bbm}
\usepackage{graphicx}
\usepackage{tabularray}

\newcommand{\secref}[1]{Section~\ref{#1}}
\newcommand{\figref}[1]{Figure~\ref{#1}}
\newcommand{\tabref}[1]{Table~\ref{#1}}

\newcommand{\utt}{\mathbf{u}}
\newcommand{\hidd}{\mathbf{h}}
\newcommand{\noise}{\boldsymbol{\delta}}

\DeclareMathOperator{\VectQuant}{VQ}
\newcommand{\vq}[1]{\VectQuant\left(#1\right)}

\newcommand{\enrolls}{A}
\newcommand{\trials}{D}
\newcommand{\va}{\text{VA}_S}
\newcommand{\train}{T}
\newcommand{\asv}{\text{ASV}}

\newif\ifcameraready
\camerareadytrue
\newacro{va}[VA]{Voice Anonymization}
\newacro{asv}[ASV]{Automatic Speaker Verification}
\newacro{vpc}[VPC]{VoicePrivacy Challenge}
\newacro{eer}[EER]{Equal Error Rate}
\newacro{nac}[NAC]{Neural Audio Codec}
\newacro{vq}[VQ]{Vector Quantization}
\newacro{pgd}[PGD]{Projected Gradient Descent}
\newacro{wer}[WER]{Word Error Rate}
\newacro{asr}[ASR]{Automatic Speech Recognition}
\newacro{snr}[SNR]{Signal-to-Noise Ratio}
\newacro{pesq}[PESQ]{Perceptual Evaluation of Speech Quality}
\newacro{stoi}[STOI]{Short-Time Objective Intelligibility}
\newacro{vc}[VC]{Voice Conversion}

\begin{document}

\title{%Proactive voice anonymization\\via adversarial perturbations
Preemptive defense against re-identification attacks on voice anonymization via adversarial perturbation
\ifcameraready
    \thanks{This work was supported by the COMPROMIS project (ANR22-PECY-0011) funded by a French government grant managed by the Agence Nationale de la Recherche under the France 2030 program.}
\else
    \thanks{This work was supported by \texttt{[text redacted for anonymity in double-blind review]}.}
\fi
}

\ifcameraready
    \author{\IEEEauthorblockN{
    Michele Panariello,
    Yibo Bai,
    Massimiliano Todisco,
    Nicholas Evans}
    \IEEEauthorblockA{\textit{Audio Security and Privacy group} \\
    \textit{EURECOM, France}\\
    \{firstname.lastname\}@eurecom.fr}
}
\else
    \author{
  \IEEEauthorblockN{Anonymous Authors}
  \IEEEauthorblockA{
    \textit{Anonymous Institution} \\
    City, Country \\
    anonymous@institution.example
  }
}
\fi

\maketitle

\begin{abstract}
% We find yet another way to promote that ``Strong defenses start with strong attacks'', except somehow it's anonymization this time? Wtf is going on?
% voice anonymization

% % overestimation has shown that privacy comes from ASV
% Voice anonymization (VA) is evaluated by an attacker's capability to re-identify speakers in anonymized utterances via automatic speaker verification (ASV), measured in Equal Error Rate (EER).
% To perform effective re-identification, the attacker needs to use \textit{unprotected} utterances of the speakers to identify as enrollment data.
% % Previous work has shown that part of the privacy protection granted by VA is due to suboptimal functioning of the attacker's ASV system.
% In this paper, we present a new \textit{proactive voice anonymization} paradigm: we protect the utterances potentially stolen by the attacker with adversarial noise, rendering them unusable as enrollments, thus preventing the attack altogether and enhancing privacy protection.
% With this approach, we are able to improve a VA system's protection level from $14\%$ EER up to nearly $50\%$ (perfect privacy).
% The proposed technique works independently of the ASV system used by the attacker, can be applied to any VA system, and does not significantly impact the quality of the utterances protected with the adversarial noise.
% % we propose a new paradigm to induce this - prevent attack from working
% % enhance privacy of any VA system up to perfect privacy without impacting on the quality
Voice anonymization (VA) is used to conceal the voice identities in speech data.
Performance is usually estimated using automatic speaker verification (ASV) as a proxy to judge the capability of an attacker to re-identify speakers after anonymization.
The defender anonymizes test utterances; the attacker compares them to equally anonymized reference utterances to infer voice identity, with degraded ASV performance indicating successful anonymization.
Thus, the defender's protection is one-sided and only applied to test utterances via VA.
Though unexplored so far, there is an opportunity to enhance anonymization by protecting reference utterances too.
We present a new, preemptive VA paradigm: reference utterances are protected using adversarial noise to degrade their potential to infer voice identity.
Preemptive protection does not degrade the perceived quality of reference utterances and is independent of the specific ASV system used for identity inference.
By combining preemptive protection and regular test-side VA, ASV equal error rates can be increased from $14\%$ to $45\%$ (near-perfect privacy).
%, and can be combined with VA approaches to further enhance privacy protection.
\end{abstract}


\section{Introduction}
\ac{va}~\cite{tomashenkoIntroducingVoicePrivacyInitiative2020} is the task of concealing the speaker's voice identity from a speech signal while preserving other linguistic and paralinguistic attributes.
% The field has recently seen considerable growth thanks to
The \ac{vpc}~\cite{vpc2020, vpc2022, vpc2024}, now in its fourth edition~\cite{vpc2026_eval_plan}, has promoted and fostered progress in the \ac{va} research community.
% provides a formal description of the task and its evaluation.
In the \ac{vpc}, the \ac{va} task is formulated as a game between a defender and an attacker:
given a set of speech utterances, the defender anonymizes them, and the attacker subsequently attempts to recognize the original voice identities.

% the fact that the literature is now split between defense and attack
Early research mostly focused on the defender side~\cite{the_one_and_only_fang, B2, srivastava20_interspeech}, while
the attack paradigms experienced a relatively slow evolution. Since the first \ac{vpc} edition~\cite{vpc2020}, the attack consists in
% the privacy adversary
using an \ac{asv} system to
% assign an identity to
identify the speaker within the anonymized utterances,
which are used as trials.
The attacker compares them to a set of reference utterances of candidate speakers they attempt to re-identify, which are used as enrollment utterances.
% they are compared against reference enrollment utterances of the speakers to re-identify, which the attacker is assumed to possess.
The higher the \ac{eer} of the \ac{asv} system, the more difficult it is for the attacker to infer the original speaker's identity, resulting in a greater degree of privacy protection.
Subsequent editions have refined this idea by expanding upon the threat model~\cite{vpc2022, vpc2024}: namely, if the attacker has knowledge of what \ac{va} system was used to anonymize the trials, they can use it to anonymize their own enrollment utterances and the training data of their \ac{asv} system, adapting to that \ac{va} system and thus improving it.
This remains the standard approach to privacy protection  evaluation in the \ac{vpc} and in the field of~\ac{va} in general.
% The adversary is assumed to have at their disposal a pool of unprotected (i.e., non-anonymized) utterances that can be used as enrollment data for their \ac{asv} system; this is known as an \textit{ignorant} attacker. Additionally, a \textit{lazy-informed} attacker is assumed to know which \ac{va} system was used to anonymize the protected data, and is therefore able to anonymize their own enrollment data to reduce mismatch with the anonymized test data.
% The second \ac{vpc} edition~\cite{vpc2022} expanded upon the threat model by assuming that the privacy adversary can train their \ac{asv} system on anonymized data, resulting in a more powerful attack: this is known as a \textit{semi-informed} attacker.

More recently, the VoicePrivacy Attacker Challenge~\cite{attacker_challenge} has promoted attack-focused research, with novel attack strategies~\cite{bakari25_nontimbral, thebaud25_spsc} and privacy evaluation metrics~\cite{legally_validated_eval, chandra2026_and_other_3000_authors} being proposed to this day.
Notably,~\cite{overestimation} introduced the idea that the \ac{eer} used as an estimate of privacy protection is influenced by two factors: \textit{(i)} how well the \ac{va} system used by the defender can suppress speaker-specific information in the signal, and \textit{(ii)} how ineffective the \ac{asv} system used by the attacker is at detecting residual speaker-specific cues that remain in anonymized utterances.
In other words, an estimate of seemingly-strong privacy protection could be due to an \textit{ineffective} attacker \ac{asv} system rather than an \textit{effective}  defender \ac{va} system.
The authors name this phenomenon \textit{privacy overestimation}
% since the attacker can potentially reduce privacy protection by improving their \ac{asv} system.
% the attacker has a certain degree of control over the privacy protection level: they can potentially reduce that protection by improving their \ac{asv} system.
and identify it as a real risk 
% The risk of privacy overestimation is very real
because, in a research setting, the evaluation of a \ac{va} system requires the developer to play both the role of the defender and the attacker:
if the developer unknowingly uses an ineffective attack, their privacy evaluation will be overly optimistic, resulting in a false sense of protection.

% if the \ac{eer} reported from that procedure relies on an ineffective attack, a real attacker 
% Overestimation can then occur more easily and unbeknownst to the developer, 
% %the reported \acp{eer} can be overly optimistic, 
% resulting in a false sense of protection.
% ramble about privacy overestimation to kickstart things

% use the word lower bound or upper bound somewhere
In this paper, we propose a technique that \textit{purposefully} exploits the component of privacy overestimation stemming from the attacker's \ac{asv} system:
we reinforce privacy protection by
% maximizing both \textit{\ac{va} effectiveness} and \textit{\ac{asv} ineffectiveness}.
maximizing \textit{\ac{asv} ineffectiveness} while keeping \textit{\ac{va} effectiveness}.
We use adversarial noise to disrupt the generation of anonymized enrollment data by the attacker.
% We use adversarial noise to perturb the attacker's use of a \ac{va} system to generate anonymized enrollment data.
The consequent mismatch between enrollment and trial data then results in a less effective attack, independently of the attacker \ac{asv} system.
% This weakens their \ac{asv} system, regardless of what deep neural network it uses as backbone.
Our contributions are as follows:
\textit{(i)} we describe the first preemptive defense against \ac{va} re-identification attacks using adversarial perturbations;
\textit{(ii)} we show that the proposed technique reinforces the privacy protection of a given \ac{va} system against \textit{any} attacker \ac{asv} system, often achieving the ``perfect privacy'' goal of 50\% \ac{eer};
\textit{(iii)} we show that the improvements to privacy protection persist even in the case in which the attacker is aware of the adversarial perturbation and attempts to remove it with purification methods~\cite{yibo_dap}.
% \begin{itemize}
%     \item we describe the first proactive defense against \ac{va} attacks using adversarial perturbations;
%     \item we show that the proposed technique reinforces the privacy protection of a given \ac{va} system against \textit{any} attacker \ac{asv} system, often achieving the ``perfect privacy'' goal of 50\% \ac{eer};
%     \item we find that the improvements to privacy protection persist even in the case in which the attacker is aware of the presence of the adversarial perturbation and attempts to remove it with purification methods~\cite{yibo_dap}.
% \end{itemize}

\begin{figure*}
    \centering
    \includegraphics[width=0.75\linewidth, trim={0 3.7cm 0 1.3cm}, clip]{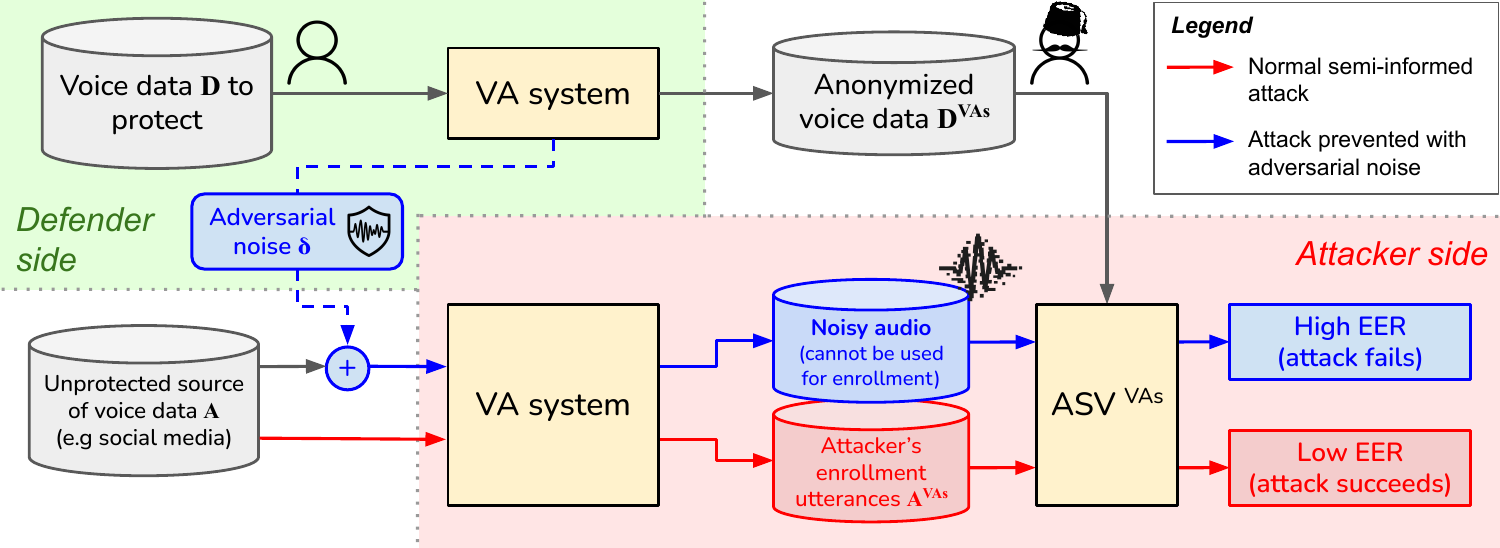}
    \caption{Overview of the proposed preemptive voice anonymization technique.}
    \label{fig:overview}
\end{figure*}

\section{Related work}
\subsection{Voice anonymization systems}
% maybe have one subsection for B4 and B5
% The usual. Describe B4 and B5. Describe semi-informed attack.
State-of-the-art \ac{va} systems function similarly to \ac{vc} systems: in practice, \ac{va} is carried out by converting the voice of the original speaker to that of a different \textit{pseudo-speaker}. In this paper, we use two baseline systems from the 2024 and 2026 editions of the \ac{vpc}: B4~\cite{B4} and B5~\cite{B5}. We choose them as they were the most competitive in terms of privacy protection in the 2024 edition.

B4 performs anonymization by extracting \textit{semantic tokens} from the utterance to anonymize with a HuBERT-based~\cite{hubert} semantic encoder.
A \ac{nac}~\cite{encodec} encoder extracts \textit{acoustic tokens} from an utterance collected from the pseudo-speaker. Acoustic and semantic tokens are combined using an autoregressive transformer to generate a new set of acoustic tokens that encode the content of the original utterance, but the voice identity of the pseudo-speaker. These are then decoded to a waveform using the \ac{nac} decoder.

B5 extracts feature vectors from the input speech with a wav2vec 2.0 backbone~\cite{wav2vec} and discretizes them with a \ac{vq} layer~\cite{neural_discrete_representation_learning}. The result is fed into a HiFi-GAN~\cite{hifigan} to synthesize the output waveform. Additionally, the HiFi-GAN is conditioned on a one-hot representation of the pseudo-speaker.

% say why you chose them: best performing + both operate on discrete units
 % We choose them as they were the best performing in terms on privacy protection in the 2024 edition of the \ac{vpc}.
% We choose to test the proposed technique on B4 and B5 for two reasons. First, they were the most competitive baselines in terms of privacy preservation in the 2024 \ac{vpc}.
% Second, both systems
Both B4 and B5 operate upon discrete representations of the input utterance: semantic tokens in B4 and \ac{vq} codewords in B5. This shared structure enables the formulation of an optimization objective for adversarial noise generation that can be applied to both systems. Further details are provided in \secref{sec:proposed_approach}.
% Second, both systems involve the computation of a tokenized representation of the input utterance, in the form of the semantic tokens and the \ac{vq} layer codewords in the case of B4 and B5 respectively. This commonality is practical in designing a single optimization objective for the adversarial noise generation routine that can be applied to both systems. See \secref{sec:proposed_approach} for further details.

\subsection{Privacy evaluation via automatic speaker verification}
\label{sec:related_works_semi-info_atk}
As per the \ac{vpc} evaluation protocol, the defender uses anonymization system $\va$ to convert unprotected utterances $\trials$ into the anonymized utterances $\trials^{\va}$.
The attacker has access to unprotected reference utterances $\enrolls$ belonging to the speakers they intend to re-identify within $\trials^{\va}$. To that end, they train an \ac{asv} system on a further dataset $\train$.
% The attacker attempts to re-identify certain speakers within $\trials^{\va}$; they are assumed to have access to unprotected reference utterances $\enrolls$ of those speakers, which are used as enrollment.
% the speakers they intend to re-identify within.
% Additionally, they possess a further dataset $\train$ that they can use to train their \ac{asv} system.
% Utterances $\enrolls$ are used as enrollment data to their \ac{asv} system.
% Since the 2022 edition of the \ac{vpc},
The \ac{vpc} privacy evaluation protocol assumes a \textit{semi-informed} attacker~\cite{vpc2024}: under this worst-case scenario, the attacker knows which system $\va$ was used by the defender to transform $\trials$ into $\trials^{\va}$. They are able to use the same system to anonymize the enrollment utterances $\enrolls$ and training data $\train$, producing $\enrolls^{\va}$ and $\train^{\va}$, respectively. They use the latter to train a speaker verification system $\asv^{\va}$ specifically adapted to $\va$. They then attempt the re-identification attack by using $\trials^{\va}$ as trials and $\enrolls^{\va}$ as enrollment utterances.
% Since the the 2022 edition of the \ac{vpc}, the standard privacy protection benchmark requires the attacker to be \textit{semi-informed}~\cite{vpc2024}: this is a worst-case scenario in which the attacker knows which \ac{va} system was used by the defender to anonymize $\trials$ into $\trials^{\va}$. The attacker can then adapt their \ac{asv} system by using the same \ac{va} system to anonymize enrollment utterances $\enrolls$ into $\enrolls^{\va}$ and training data $\train$ into $\train^{\va}$, obtaining an adapted system $\asv^{\va}$.
% their enrollment utterances and the anonymized test data by anonymizing with the same \ac{va} system both the enrollment utterances and the training data of the attacking \ac{asv} system.

This approach constitutes the strongest attacks known to date~\cite{vpc2026_eval_plan}, and is therefore considered to provide a lower bound for the privacy protection provided by a \ac{va} system (with some notable exceptions~\cite{overestimation}). Preemptive anonymization is designed to hinder a semi-informed attack, and therefore raise this lower bound.

\subsection{Adversarial perturbations in audio security and privacy}
To the best of our knowledge, this is the first work to investigate  adversarial perturbations as a means of enhancing the privacy protection of \ac{va} systems. The closest line of related research studies adversarial perturbations to defend against voice cloning~\cite{voiceguard, liping_adv1, liping_adv2, park25_spsc}: these methods focus on generating noise that disrupts the output embeddings of one or more speaker encoders, preventing the use of protected utterances in a voice cloning system.
Our approach is different in that we generate noise that degrades at the waveform level the reference enrollment utterances used by the attacker. Comparing them to the anonymized trials produces unreliable similarity scores and causes the re-identification attack to fail.
% that prevents the attacker from using an utterance as enrollment for $\asv^{\va}$. In practice, this is done by making it impossible for the attacker to anonymize said utterance; without anonymized enrollment utterances $\enrolls^{\va}$, the semi-informed attack fails.

% The noise is optimized to degrade the output of the \ac{va} system itself: if the attacker 

% that degrades the output of the \ac{va} system used to anonymize $\trials^{\va}$, so that the attacker cannot use it as enrollment data for its \ac{asv} system.

% The use of adversarial noise to disrupt \ac{asv} systems has been explored in the literature~\cite{universal_perturbations_CM, malacopula}. These methods directly target the \ac{asv} system in question, and can be less effective in black-box scenarios; in contrast, our approach renders the potential enrollment audio unusable before the \ac{asv} is even applied. As we show in Section \textcolor{red}{stocazzo}, this is effective even when the defender has no information about the attacker's \ac{asv}, in accordance with the threat model of the \ac{vpc}.

\section{Preemptive defense against re-identification attacks}
\label{sec:proposed_approach}
The semi-informed attack scenario is based on two main assumptions: \textit{(i)} the attacker somehow obtains unprotected enrollment utterances $\enrolls$ of the speakers they intend to re-identify within the anonymized data $\trials^{\va}$; and \textit{(ii)} the attacker knows which system $\va$ was used by the defender to anonymize $\trials$ into $\trials^{\va}$.
Assumption \textit{(i)} requires the existence of a source from which the attacker can retrieve unprotected speech data of the speakers of interest.
Social media provides a natural example, as users generally wish to remain recognizable and are therefore unlikely to anonymize their voices. The attacker could also gain unauthorized access to private speech data that would not normally be anonymized, such as voice messages exchanged through messaging services.
% The most obvious example is social media, where users normally want to be recognized as themselves, and therefore would not anonymize their voice.
% Alternatively, the attacker could maliciously access private data that is not normally anonymized like voice notes in messaging services.
% In designing our proactive anonymization technique, we rely on a third assumption \textbf{(iii)}: that the defender has some control over the voice data that is at risk of being stolen by the attacker. Such a risk is prevented by adding adversarial perturbations to the data.
Our proactive anonymization method introduces a third assumption, \textit{(iii)}: the defender has some control over the speech data $\enrolls$ that the attacker may potentially obtain. The defender can then protect these utterances in advance (e.g. before upload to social media) by adding adversarial perturbations, thereby preventing the attacker from using them as enrollment data.
% is aware of this risk, and are able to add adversarial perturbations to the utterances that could potentially be stolen by an attacker.
The proposed approach is illustrated in \figref{fig:overview}.

The perturbations must respect two requisites.
First, they should preserve the quality of the protected utterance;
% Remember to say why we can't attack the ASV directly
second, they cannot be optimized directly using $\asv^{\va}$ since the defender cannot know the \ac{asv} system used by the attacker.
However, the defender does know  the \ac{va} system they used to generate $D^{\va}$.
For a given utterance $\utt$, the defender can then generate an adversarial perturbation $\noise$ in white-box fashion so that passing $\utt + \noise$ through $\va$ will result in a highly degraded output: in other words, they prevent the protected signal from being anonymized.
% through the \ac{va} system results in random noise rather than an anonymized utterance.
% it is thus possible for them to generate adversarial noise that makes it unusable in a white-box fashion.
If the attacker's enrollment data $\enrolls$ is preemptively protected with these perturbations, anonymizing it into $\enrolls^{\va}$ will result in noisy data that will mislead $\asv^{\va}$, making the semi-informed attack fail.
% If the attacker steals the protected data and tries to anonymize it with the same \ac{va} system for a semi-informed attack, they will fail.
% Attempts of anonymizing the same data with a different \ac{va} system to circumvent the proactive defense would result in a suboptimal attack~\cite{overestimation}.
% Proactive protection works by making the worst-case scenario of a semi-informed attack impossible.
The attacker cannot circumvent the noise by using a different \ac{va} system: according to previous research, this would cause a distribution mismatch between and $\enrolls^{\va}$ and $\trials^{\va}$, resulting in a poor attack~\cite{overestimation}.

% The technique, explained on both B4 and B5.
Exploiting the white-box setting, the adversarial perturbation is generated by targeting the internal modules of a given \ac{va} system. We describe this process first for B5, then for B4.
% therefore, the optimization process must be tailored to each system individually. We describe it for systems B4 and B5. We start from the latter since it has the simpler topology.

\subsection{Attack on B5}
Consider an utterance $\utt \in \mathbb{R}^L$ to be anonymized. As an intermediate processing step, the utterance is encoded into a set of feature vectors using wav2vec 2.0~\cite{wav2vec}, with $\hidd_t$ being the feature vector at time frame $t$. Each $\hidd_t$ is subsequently mapped to a codeword by a \ac{vq} layer. Our goal is to find a perturbation $\noise \in \mathbb{R}^L$ which degrades the output of B5 as much as possible when added to $\utt$. To this end, we maximize the number of \textit{codeword flips}, defined as time frames for which the perturbed utterance is assigned a \ac{vq} codeword different to  the original codeword. The resulting optimization objective is:
\begin{equation}
\label{eq:true_objective}
    \max_{\noise} \frac{1}{T} \sum_{t=1}^T \mathbbm{1}\left[\ \vq{\hidd_t}\ \neq\ \vq{\hidd_t^{\noise}}\ \right]
\end{equation}
where $\mathbbm{1}[\cdot]$ is the indicator function, $T$ is the total number of time frames, and $\hidd_t^{\noise}$ is the feature vector extracted from $\utt + \noise$ at time step $t$.
This objective cannot be optimized via gradient descent because of the non-differentiable \ac{vq} operation; however, an equivalent, differentiable surrogate objective can be formulated as:
\begin{equation}
\label{eq:surrogate_objective}
    \max_{\noise} \frac{1}{T} \sum_{t=1}^{T} \left|\hidd_t - \hidd_t^{\noise}\right|_1
\end{equation}
whereby we attempt to push $\hidd_t$ outside of its original Voronoi region~\cite{neural_discrete_representation_learning} to produce a codeword flip.
We found the L1 distance to produce more consistent results than the more conventional mean squared error, since it produces smaller loss values and therefore more stable gradients.

Following other works~\cite{psychoacoustic_masking}, we limit the impact of perturbations upon audio quality by constraining $\noise$ to a ball of radius $\varepsilon$.
% To limit the impact of the perturbation on the audio quality, $\noise$ is constrained to a ball of radius $\varepsilon$.
This value can be tuned to adjust the privacy-utility tradeoff, with larger $\varepsilon$ resulting in stronger preemptive protection but potentially greater degradation to quality.
%a potentially more degraded utterance. 

In B5, if inference is done in batches, utterances are normally zero-padded to the right. Preliminary experiments showed that this padding can reduce the effectiveness of adversarial noise. To account for this, we also apply right zero-padding to $\utt$ while optimizing \eqref{eq:surrogate_objective}, making $\noise$ robust to the padding introduced at inference time. The padding expands $\utt$ to a different random length $L'>L$ at every iteration step.
Adversarial noise is not applied to the padded area ($\noise$ is always of length $L$). Code of the optimization routine is available at \url{https://github.com/eurecom-asp/preemptive_VA_defense}.

\subsection{Attack on B4}
B4 includes a similar quantization step whereby the input utterance $\utt$ is tokenized into semantic tokens using a semantic encoder. An optimization objective similar to \eqref{eq:true_objective} can therefore be applied by replacing the \ac{vq} layer with the tokenization operation.
In practice, we again optimize~\eqref{eq:surrogate_objective} instead, where $\hidd_t$ are now the pre-tokenization feature vectors extracted using the semantic encoder.
Since B4 does not support batched inference, no padding is ever applied to the inputs; therefore, $\noise$ does not need robustness to zero-padding.

Note that the surrogate objective~\eqref{eq:surrogate_objective} is generic enough to be applied to any \ac{va} system that extracts time-ordered embeddings from the input utterance.

\begin{table}[t]
\centering
\caption{Privacy preservation results on \textit{libri\_test}. EER (\%) $\uparrow$ is reported for the \ac{vpc} 2024 and \ac{vpc} 2026 attackers.}
\label{tab:privacy_results}
\begin{tblr}{
    colspec = {Q[c,m] Q[l,m] Q[c,m] Q[c,m]},
    vlines,
    hline{1,2,7,12} = {1.2pt},
    hline{3-6,8-11} = {0.4pt},
    row{1} = {font=\bfseries},
    cell{2,7}{2-4} = {bg=gray9},
    column{1} = {font=\bfseries},
    % column{2} = {font=\bfseries},
}
\SetCell[c=2]{c} System & & VPC'24 attacker & VPC'26 attacker \\
\SetCell[r=5]{c} B4
    & Baseline (no noise)       & 30.26 & 14.33 \\
    & $+\noise$ noise, $\varepsilon=0.05$  & 36.63 & 19.99 \\
    & $+\noise$ noise, $\varepsilon=0.1$   & 41.18 & 30.59 \\
    & $+\noise$ noise, $\varepsilon=0.3$   & 46.67 & 44.51 \\
    & $+\noise$ noise, unbounded         & 47.51 & 45.14 \\
\SetCell[r=5]{c} B5
    & Baseline (no noise)       & 34.34 & 21.28 \\
    & $+\noise$ noise, $\varepsilon=0.05$  & 35.75 & 26.26 \\
    & $+\noise$ noise, $\varepsilon=0.1$   & 41.93 & 37.30 \\
    & $+\noise$ noise, $\varepsilon=0.3$   & 45.15 & 47.36 \\
    & $+\noise$ noise, unbounded         & 46.33 & 46.64 \\
\end{tblr}
\end{table}

\section{Experimental setup}
\subsection{Optimization parameters}
We optimize $\noise$ for each utterance with \ac{pgd}~\cite{pgd}.
Gradients are computed with Adam~\cite{adam}. 
Optimization is carried out until one of the following conditions are met: \textit{(a)} we reach a maximum of 700 iterations; and \textit{(b)} the value of \eqref{eq:true_objective} is at least $0.95$,
% for the current utterance,
implying that a codeword flip has occurred for $95\%$ of the noisy feature vectors $\hidd_t$.
% implying that $95\%$ of the noisy embeddings have changed their assigned codeword with respect to the original ones.
The Adam learning rate is set to $10^{-4}$ for the first $300$ iterations and $10^{-3}$ thereafter. The maximum duration of the random padding $L'$ is set to 3 seconds.
% in $95\%$ of the cases, $\hidd_t^{\noise}$ as been assigned to a different codeword than the original 
% 3 conditions of epsilon
To explore the tradeoff between privacy protection and quality preservation, we perform experiments with $\varepsilon$ values of $0.3$, $0.1$, $0.05$. We also report results for unbounded noise, equivalent to $\varepsilon=\infty$.

\subsection{Privacy preservation metrics}
To show that preemptive anonymization works independently of the attacker \ac{asv} system, we perform privacy evaluation using the attackers of both the 2024 and 2026 \ac{vpc} editions. The former is based on ECAPA-TDNN~\cite{the_one_and_only_ecapa}, while the latter additionally uses a WavLM front-end~\cite{arefeen2026dastdualstreamvoiceanonymization}.
% As per the standard \ac{vpc} evaluation protocol, the privacy preservation metric is the \ac{asv} \ac{eer}.

Privacy evaluation builds upon the standard \ac{vpc} protocols described in~\cite{vpc2024, vpc2026_eval_plan}. The privacy evaluation metric is the $\asv^{\va}$ \ac{eer}.
In our primary evaluation scenario, the defender applies adversarial perturbations to the enrollment partition of the \textit{libri\_test} dataset. Following the usual \ac{vpc}-defined semi-informed attack,
%protocol defined by the \ac{vpc}, the attacker anonymizes 
these protected utterances are further anonymized and used 
%the resulting speech 
as enrollment data $\enrolls^{\va}$. The corresponding $\asv^{\va}$ is trained using an anonymized version of \textit{libri\_train-360}.

Additionally, we consider a secondary privacy evaluation scenario involving a stronger attacker who is aware that the enrollment utterances have been protected with adversarial perturbations and who attempts to remove them through adversarial noise purification. We evaluate three purification methods. The first, proposed in~\cite{chang2021defending}, adds Gaussian noise to protected utterance; we consider standard deviations of $0.05$ and $0.01$. AudioPure~\cite{wu2023defending} relies on an unconditional diffusion model trained using clean speech; we apply a single purification step. Finally, DAP~\cite{yibo_dap} uses a diffusion model trained using speech data containing adversarial perturbations.% against ECAPA-TDNN~\cite{the_one_and_only_ecapa}.
% We also evaluate an additional scenario in which the attacker has an even higher degree of knowledge: they are aware that adversarial noise was applied to the enrollment utterances, and attempt to remove it via adversarial noise purification.
% Three purification methods are employed in this scenario. The first one~\cite{chang2021defending} consists in adding Gaussian noise to the protected utterance; we experiment with standard deviation set to $0.05$ and $0.01$. AudioPure~\cite{wu2023defending} uses an unconditional diffusion model trained on clean speech; we use a single purification step. DAP~\cite{yibo_dap} uses a diffusion model trained on speech with adversarial perturbations.

% \begin{itemize}
%     \item ''normal'' EER
%     \item SRD?
%     \item scenario where the attacker is aware and uses purification
% \end{itemize}

\subsection{Quality preservation metrics}
Following the \ac{vpc} protocol, our primary quality preservation metric is the \ac{wer} of the same \ac{asr} system used in the 2024 and 2026 \ac{vpc}.
% We use 
% %that uded in 
% the \ac{asr} system of the 2024 and 2026 \acp{vpc}~\cite{vpc2024, vpc2026_eval_plan}.
% from the 2024 and 2026 editions of the Challenge~\cite{vpc2024, vpc2026_eval_plan}.
Unlike the standard \ac{vpc} setting, for which the \ac{wer} is measured using anonymized trial utterances $\trials^{\va}$, our objective is to preserve the intelligibility of enrollment utterances $\enrolls$ after adversarial perturbation but before anonymization. We therefore compute the \ac{wer} using  perturbed enrollment utterances $\enrolls + \noise$ and compare it to the \ac{wer} obtained by the same \ac{asr} system when applied to original enrollment utterances $\enrolls$;
the smaller the gap, the better the quality preservation.
% suggests that the presence of adversarial noise does not substantially affect the ASR system.
% Unlink the \ac{vpc} setting, in which \ac{wer} is computed on the anonymized trial utterances $\trials^{\va}$, our goal is to preserve the quality of enrollment utterances $\enrolls$ that are protected with adversarial noise, but not yet anonymized with a \ac{va} system. Thus, for assessment, we report the \ac{wer} on the enrollment utterances protected with the noise $\noise$, and compare that value to the \ac{wer} obtained by the same \ac{asr} system on the original enrollment utterances.
% $\Delta$\ac{wer}, which is defined as the increase in \ac{wer} between the original enrollment utterances and the enrollments utterances protected with adversarial noise.

\ac{wer} only captures intelligibility from the perspective of an \ac{asr} system. For more comprehensive quality assessment, we also compute the following secondary quality metrics between original and noise-protected enrollment utterances: \ac{snr} expressed in dB, \ac{pesq}~\cite{pesq} and \ac{stoi}~\cite{stoi}.

\begin{table}[t]
\centering
\caption{Quality preservation results on \textit{libri\_test\_enrolls}. The \ac{wer} (\%) of the original data is $1.85$.}
\label{tab:quality_results}
\resizebox{\columnwidth}{!}{%
\begin{tblr}{
    colspec = {Q[c,m] Q[l,m] Q[c,m] Q[c,m] Q[c,m] Q[c,m]},
    vlines,
    hline{1,2,6,10} = {1.2pt},
    hline{3-5,7-9} = {0.4pt},
    row{1} = {font=\bfseries},
    column{1} = {font=\bfseries},
    % column{2} = {font=\bfseries},
}
\SetCell[c=2]{c} System & & \ac{wer} $\downarrow$ & \ac{snr}\textsubscript{dB} $\uparrow$ & \ac{pesq} $\uparrow$ & \ac{stoi} $\uparrow$ \\
\SetCell[r=4]{c} B4
    & $+\noise$ noise, $\varepsilon=0.05$ & 2.05  & 49.87 & 4.48 & 0.99 \\
    & $+\noise$ noise, $\varepsilon=0.1$  & 2.39  & 43.90 & 4.30 & 0.99 \\
    & $+\noise$ noise, $\varepsilon=0.3$  & 7.01  & 34.38 & 3.51 & 0.99 \\
    & $+\noise$ noise, unbounded        & 29.58 & 26.70 & 2.55 & 0.97 \\
\SetCell[r=4]{c} B5
    & $+\noise$ noise, $\varepsilon=0.05$ & 1.88 & 49.88 & 4.49 & 0.99 \\
    & $+\noise$ noise, $\varepsilon=0.1$  & 1.97 & 43.89 & 4.30 & 0.99 \\
    & $+\noise$ noise, $\varepsilon=0.3$  & 2.19 & 34.39 & 3.53 & 0.99 \\
    & $+\noise$ noise, unbounded        & 2.59 & 27.32 & 2.61 & 0.97 \\
\end{tblr}}
\end{table}

\section{Results}
\subsection{Privacy protection}
% We report the privacy protection results in \tabref{tab:privacy_results}. In the case of the \ac{vpc} 2024 attacker, B4 and B5 achieve a baseline \ac{eer} of around $30\%$ and $34\%$ respectively. By applying the adversarial noise for proactive protection in the unbounded setting ($\varepsilon=\infty$), we are able to increase these values to $47\%$ and $46\%$, nearing the ideal perfect privacy \ac{eer} value of $50\%$.
Privacy results are reported in \tabref{tab:privacy_results}. With the \ac{vpc} 2024 attacker \ac{asv} system, B4 and B5 obtain baseline \acp{eer} of approximately $30\%$ and $34\%$, respectively. Preemptive protection with unbounded adversarial perturbations ($\varepsilon=\infty$) increases \acp{eer} to $47\%$ for B4 and $46\%$ for B5, approaching the perfect privacy level of $50\%$.
The \ac{vpc} 2026 attacker is substantially stronger, and reduces the baseline \ac{eer} to $14\%$ for B4 and $21\%$ for B5. Nevertheless, the proposed adversarial perturbations increase the \acp{eer} to approximately $45\%$ and $46\%$, respectively, again bringing performance close to the perfect privacy level.
These results suggest that proactive protection remains effective across different attacker \ac{asv} systems. This is because adversarial perturbation targets the \ac{va} system itself, rather than a specific \ac{asv} model. When objective~\eqref{eq:true_objective} reaches a sufficiently high value, casual listening tests reveal that the anonymized enrollment utterance becomes largely unintelligible,\footnote{Audio samples are provided in the supplementary material.} obscuring identifiable information regardless of the \ac{asv} system.
% This shows that proactive protection works for any \ac{asv} system the attacker may employ. This is because the adversarial noise targets the \ac{va} system itself rather than the \ac{asv} system, and erases speaker-specific information from the utterance before \ac{asv} is even applied. Indeed, in the case where objective~\eqref{eq:true_objective} reaches a reasonably high value, the anonymized enrolled utterance becomes completely unintelligible, suggesting that there might be no personally-identifiable information left for any \ac{asv} system to use.
% Audio samples are provided in the supplementary material.

The method remains similarly effective when the perturbation magnitude is bounded by $\varepsilon=0.3$. Across both B4 and B5, and for both the 2024 and 2026 attackers, the \acp{eer} again reach approximately $46\%$.
% The method remains equally effective with a magnitude bound of $\varepsilon=0.3$. For both B4 and B5, and for both 2024 and 2026 attackers, the \acp{eer} are again pushed to around $46\%$. Indeed, we observe that the stopping condition of achieving a value of $0.95$ on~\eqref{eq:true_objective} is often met.
Privacy gains drop when the perturbation bound is reduced to $\varepsilon=0.1$. For the 2024 attacker, the \acp{eer} decrease to approximately $41\%$ for both B4 and B5. The stronger 2026 attacker lowers them further, to $30\%$ and $37\%$, respectively. Even so, these values remain approximately twice as high as the corresponding baseline \acp{eer}. Finally, a smaller bound of $\varepsilon=0.05$ still provides modest privacy improvements.

\subsection{Quality preservation}
Quality preservation results are shown in \tabref{tab:quality_results}. In general,  higher noise bounds $\varepsilon$ lead to greater impacts on quality.
% The initial \ac{wer} on the unprotected enrollment utterances is $1.85\%$.
% In the case of B5, the addition of noise increases this value by a negligible amount even in the case of \mbox{$\varepsilon=\infty$}, which results in a $2.6\%$ \ac{wer}; further decreasing the value of $\varepsilon$ has an even lesser impact. All secondary quality metrics follow a similar trend: the \ac{snr} starts at $27.3$ dB and keeps increasing, suggesting that the noise is modest compared to the actual signal; intelligibility is fully preserved, with all \ac{stoi} values nearing their upper value $1$; \ac{pesq} values are solid, with $\varepsilon=0.3$ resulting in a score of $3.5$ (with the \ac{pesq} range being $[-0.5, 4.5]$).
The baseline \ac{wer} for unprotected enrollment utterances is $1.85\%$. For B5, the addition of adversarial perturbation causes only a minor increase, even in the unbounded setting \mbox{$\varepsilon=\infty$}, for which the \ac{wer} reaches $2.6\%$. Smaller perturbation bounds $\varepsilon$ have an even smaller effect. The secondary quality metrics show a similar trend. The \ac{snr} is $27.3$~dB for the unbounded setting and increases as $\varepsilon$ decreases, indicating that adversarial perturbations remain imperceptible relative to the speech signal. Intelligibility is fully preserved; all \ac{stoi} scores are close to their maximum value of $1$. \ac{pesq} scores also remain high, ranging from approximately $3.51$ ($\varepsilon=0.3$) to $4.48$ ($\varepsilon=0.05$) for a maximum value of $4.5$.
% For example, for $\varepsilon=0.3$, \ac{pesq} scores are approximately $3.5$ in the range of $[-0.5, 4.5]$.

% Conversely, the impact of the noise generated for B4 is apparently more pronounced: setting $\varepsilon=\infty$ results in a surprisingly high \ac{wer} of $29.6\%$. The issue can be mitigated by setting $\varepsilon=0.3$ or lower, which diminishes the \ac{wer} to around $7\%$.
% This result is unexpected since all other secondary quality metrics are in line with the case of B5, suggesting that the adversarial noise itself is not stronger in magnitude; informal listening tests confirm this.
In contrast, perturbations generated for B4 have a less negligible impact on the \ac{wer}: for the unbounded setting, \mbox{$\varepsilon=\infty$}, the \ac{wer} rises unexpectedly to $29.6\%$. This degradation is substantially reduced when the perturbation is bounded, with $\varepsilon=0.3$ lowering the \ac{wer} to approximately $7\%$ and tighter bounds yielding further improvements. Despite the rise in \ac{wer}, the secondary quality metrics are comparable to those obtained for B5, indicating that the perturbation is not larger in magnitude. Informal listening tests lead to the same conclusion.
% We therefore hypothesize that the abnormally high \ac{wer} in the case of B4 might be due to the adversarial noise $\noise$ having an undesired \textit{transfer} effect on the specific \ac{asr} architecture that is being used for evaluation: that is, optimizing $\noise$ to disrupt B4 results in $\noise$ also being effective in disrupting this specific \ac{asr} system's backbone.
% We find partial confirmation of our hypothesis by running the same experiment with a different \ac{asr} system\footnote{\url{https://huggingface.co/speechbrain/asr-streaming-conformer-librispeech}} on the same data: in the case of $\varepsilon=\infty$, we obtain a much lower \ac{wer} of $6.6\%$.\footnote{Since B4 and the employed \ac{asr} do not have any internal modules in common, the reasons of this transferability are still a matter of investigation. It is worth noting that B5 is actually architecturally more similar to the \ac{asr} system in question, since both their backbones are based on some version of wav2vec --- yet, the effects of $\noise$ does not seem to transfer between them.}
%Our hypothesis is that the unusually high \ac{wer} observed for B4 is 
Further investigation suggests that
the cause is likely an unintended \textit{transfer} effect: although $\noise$ is optimized to disrupt B4, it may also interfere with the backbone of the specific \ac{asr} system used for evaluation. We find partial support for this hypothesis by repeating the experiment with a different \ac{asr} system\footnote{\url{https://huggingface.co/speechbrain/asr-streaming-conformer-librispeech}} for which, in the unbounded setting, the \ac{wer} is substantially lower at $6.6\%$.\footnote{The cause of this transfer remains unclear, since B4 and the original \ac{vpc} \ac{asr} system do not share any internal modules. Interestingly, B5 is architecturally closer to the \ac{vpc} \ac{asr} system; both use wav2vec-based backbones, yet no comparable transfer effect is observed.}

Overall, these results indicate that the quality of protected utterances is sufficiently preserved for use in other downstream tasks. In particular, the high \ac{pesq} and \ac{stoi} scores suggest that $\noise$ has only a limited perceptual impact, making the method suitable for the protection of speech shared on social media (see use case described in \secref{sec:proposed_approach}).

\begin{figure}
    \centering
    \includegraphics[width=\columnwidth]{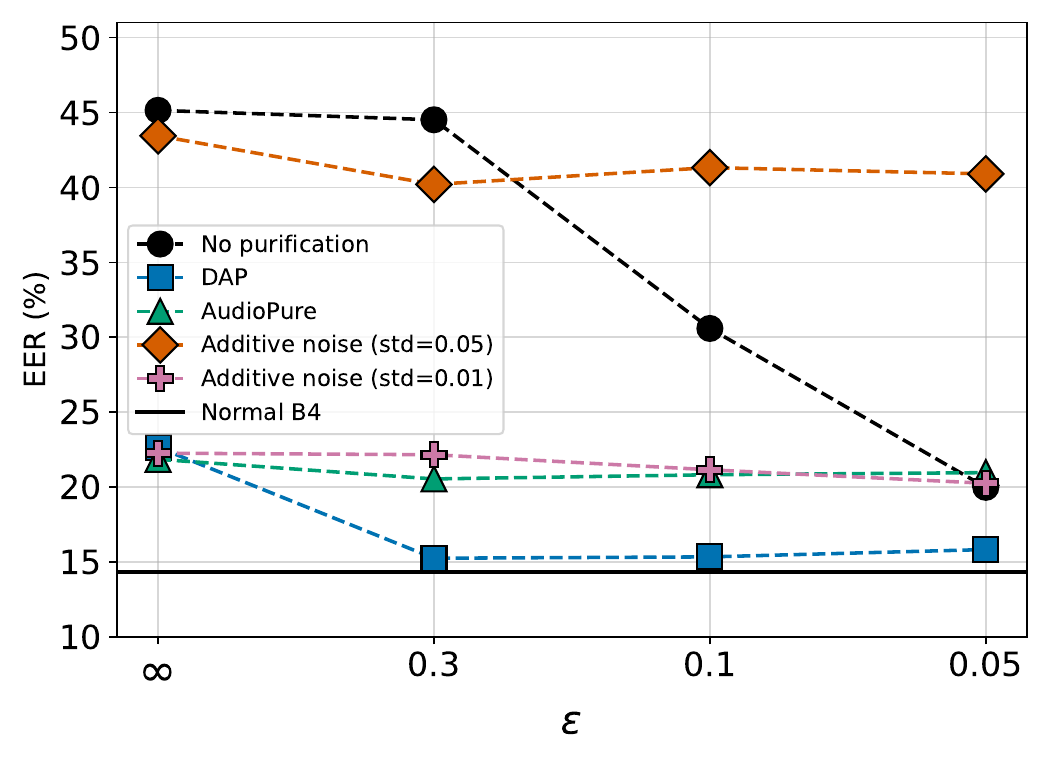}
    \vspace{-0.8cm}
    \caption{Preemptive privacy protection results when the attacker attempts to remove the adversarial noise $\noise$ from the enrollment utterances with various purification methods. The experiments are run in the case of B4 with the \ac{vpc} 2026 attacker (worst case scenario in terms of privacy protection).}
    \label{fig:purification_results}
\end{figure}

\subsection{Use of adversarial noise purification by the attacker}
\figref{fig:purification_results} presents results for the scenario in which the attacker is aware of the presence of adversarial perturbations and attempts to remove them through purification. For conciseness, we consider the most challenging defender setting: using B4, which provides weaker privacy protection than B5, against the stronger \ac{vpc} 2026 attacker.

% The emergent behavior is that most purification systems manage to remove the perturbation, but in turn add some disturbance of their own which still partially disrupts the attack. For example, AudioPure and random additive noise with standard deviation of $0.01$ converge to an \ac{eer} of $20$ to $22\%$ regardless of the value of $\varepsilon$, which is still higher than the baseline \ac{eer} of $14.3\%$. Random noise with higher standard deviations degrade the signal even more than the optimize $\noise$ itself and hinder the attack even more.
Most purification methods are successful in attenuating adversarial perturbation, but also introduce distortions which have the effect of reducing the effectiveness of the attack.
For instance, AudioPure and additive Gaussian noise with a standard deviation of $0.01$ yield \acp{eer} of between $20\%$ and $22\%$ across all values of $\varepsilon$, remaining above the baseline \ac{eer} of $14.3\%$. Gaussian noise with larger standard deviations degrades the signal even more than optimized perturbations $\noise$, resulting in an even stronger attack disruption.
DAP is the only method that almost completely removes the effect of proactive protection for $\varepsilon \leq 0.3$. 

% DAP seems to be the only system to nullify almost completely the effect of the proactive protection for values of $\varepsilon$ of $0.3$ or less.
% It could be argued that future work might involve strengthening the adversarial noise against purification, falling back into the cat-and-mouse game between defender and attacker which has occurred many times in the \ac{va} literature~\cite{panariello_thesis}.
% However, it should also be noted that, even when removed, proactive protection still has the evident benefit of \textit{increasing the attacker's cost} to carry out an effective semi-informed attack. Even in the worst-case scenario in which the attacker has knowledge of the proactive protection, they still have to invest computing resources in training and applying an effective purification technique like DAP, which requires running a diffusion model among other things.
% Therefore, the proposed proactive protection technique still acts as a considerable deterrent for an attacker.
One possible direction for future work would therefore be to make the adversarial perturbation more robust to purification, continuing the cat-and-mouse dynamic between defenders and attackers that characterizes the recent \ac{va} literature~\cite{panariello_thesis}.
Even when purification is successful, however, proactive protection still increases the cost of carrying out an effective semi-informed attack. In the worst-case scenario, where the attacker is aware of the defense, they must still devote computational resources to training and applying a suitable purification method such as DAP, which requires the use of a computationally demanding diffusion model. The proposed approach therefore remains a meaningful deterrent by raising the effort required from the attacker.

% if we use another model we get 6.671

% To show:
% \begin{itemize}
%     \item two bar plots, either: grouped by ASV showing lines for B4 and B5, or grouped by systems showing lines for VPC24 and VPC26 attacker. In one of them will be the extra line for the purificaton case. the metrics here will be EER and WER (the two official VPC metrics)
%     \item additional quality metrics table (SNR, STOI, PESQ)
%     \item one more table for SRD?
% \end{itemize}

\section{Conclusions}
We presented the first approach to voice anonymization in which a defender acts preemptively to hinder potential re-identification attacks. Adversarial perturbations are used to protect clear speech data that might be obtained by an attacker, thereby making these utterances unsuitable for use as enrollment data in a semi-informed attack.

Our results show that the proposed method greatly enhances the privacy protection of a given voice anonymization system independently of the attacker ASV system.  Speech quality is preserved, thereby safeguarding the use of treated enrollment data for other downstream tasks.
% also applicable regardless of what else you are anonymizing
Adversarial noise is applied to enrollment utterances retroactively, rather than to anonymized trial utterances; thus, the approach is complementary to regular voice anonymization, and enhances the protection of trial utterances that were anonymized even before the adversarial perturbations were added to the enrollment utterances. 
%, our approach can be used retroactively to enhance the privacy protection of data that was anonymized even before the noise was applied.%, as long as the voice anonymization is done with the same system.

% limitation: the adv noise can be improved
% A limitation of the present work is that the
Our results demonstrate the potential of preemptive anonymization.
Future work should improve adversarial noise generation, 
%procedure could be further improved, 
either by increasing robustness to purification or by further reducing perceptual impacts, for instance through psychoacoustic masking~\cite{psychoacoustic_masking, schonherr2019adversarial}. 
%Reported results nonetheless demonstrate the potential 
%The aim of this study was to validate the general idea and feasibility of 
%for proactive anonymization; refinements to the noise generation procedure are hence left for future work.

\newpage
% \section*{Acknowledgment}
% That random guy at the gym a couple of months ago who tried to show me how to do pull-ups properly. It didn't help much to be honest, but it was nice of him.

% Also, this line from the original template might be relevant:
% \textit{Put sponsor acknowledgments in the unnumbered footnote on the first page.}

% So we probably don't need this section
\section{AI-generated content disclosure}
LLM-based agents were used to assist in the development of some parts of the source code. LLM-based text generation was used to correct and rephrase part of the manuscript. The research question, proposed methodology and evaluation setup, as well as the manuscript structure, were formulated exclusively by the authors, without the assistance of AI.

\bibliographystyle{IEEEtran}
\bibliography{IEEEabrv,bib_file}

\end{document}